# Probing redox potential for Iron sulfur clusters in photosystem I

Fedaa Ali[1], Medhat W.Shafaa[1], Muhamed Amin[2,3]*

†Medical biophysics division, Department of Physics, Faculty of Science, Helwan university, Cairo, Egypt.

‡ [a] Department of Sciences, University College Groningen, University of Groningen, Hoendiepskade 23/24, 9718 BG Groningen, Netherlands

Universiteit Groningen Biomolecular Sciences and Biotechnology Institute, University of Groningen, Groningen, Netherlands

*E-mail: m.a.a.amin@rug.nl

**Abstract.** Photosystem I is a light-driven electron transfer device. Available X-ray crystal structure from Thermosynechococcus elongatus, showed that electron transfer pathways consist of two nearly symmetric branches of cofactors converging at the first iron sulfur cluster $F_X$, which is followed by two terminal iron sulfur clusters $F_A$ and $F_B$. Experiments have shown that $F_x$ has lower oxidation potential than $F_A$ and $F_B$, which facilitate the electron transfer reaction. Here, we use Density Functional Theory and Multi-Conformer Continuum Electrostatics to explain the differences in the midpoint $E_m$ potentials of the $F_x$, $F_A$ and $F_B$ clusters. Our calculations show that $F_x$ has the lowest oxidation potential compared to $F_A$ and $F_B$ due strong pair-wise electrostatic interactions with surrounding residues. These interactions are shown to dominated by the bridging sulfurs and cysteine ligands, which may be attributed to the shorter average bond distances between the oxidized Fe ion and ligating sulfurs for $F_X$ compared to $F_A$ and $F_B$. Moreover, the electrostatic repulsion between the 4Fe-4S clusters and the positive potential of the backbone atoms is least for $F_X$ compared to both of $F_A$ and $F_B.$ These results agree with the experimental measurements from the redox titrations of low-temperature EPR signals and of room temperature recombination kinetics.

**Keywords.** Photosystem I . Iron-sulfur cluster . Continuum electrostatics . Broken symmetry DFT. Electron transfer . MCCE

## Introduction.

Photosynthesis process is the process that guarantee the existence of our life. In photosynthesis, the solar energy is harvested by pigments associated with the photosynthetic machinery and stored as energy rich compounds[1]. Initial energy conversion reactions take place in special protein complexes known as Type I and Type II reaction centers[2]. Which are classified according to the type of terminal electron acceptor used, iron-sulfur clusters (Fe-S) and mobile quinine for type I and type II, respectively[2–7]. Photosystem I (PS I) is Type I reaction center found in the thylakoid membranes of chloroplasts and cyanobacteria[6,8]. PS I is very interesting electron transfer machine which converts the solar energy to a reducing power with a quantum yield close to $1$ [9–11]. It, mainly, mediates the transfer of electrons

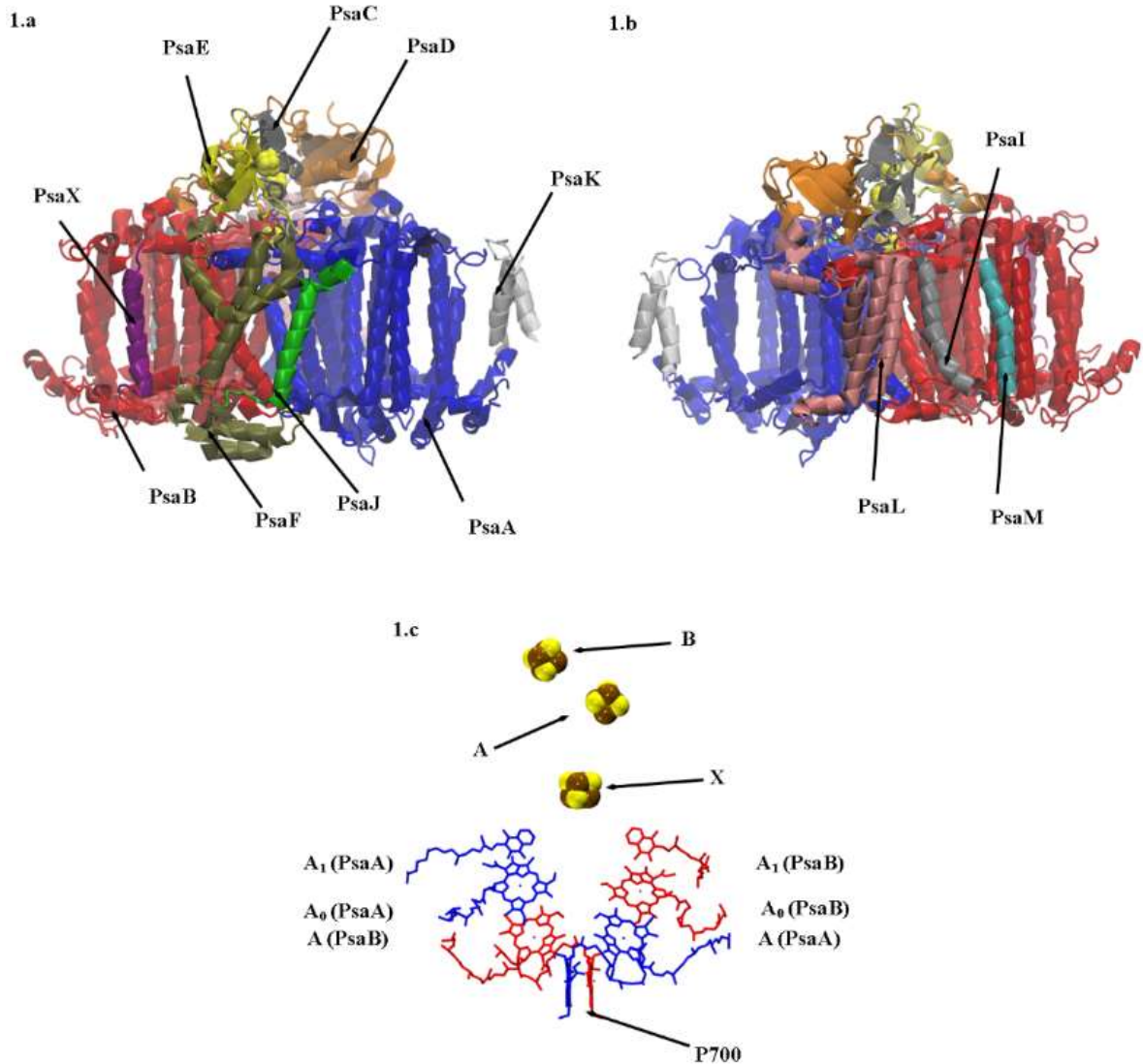

**Figure 1.** [PDB code 1JB0[12]] : 12 protein-subunits in the polypeptide structure of Cyanobacterial PS I monomer viewed perpendicular of the plane of the thylakoid membranes. 1.a: Front view and 1.b: Back view while 1.c: Shows the electron transfer chains ETCs in PS I, where P700 is primary electron donor (Chl a dimer), primary electron acceptors A /$A_0$ (Chl a molecules), secondary electron acceptor $A_1$ ( Phylloquinone molecule PQN), tertiary electron acceptor X ($F_X$) and terminal electron acceptor A ($F_A$) and B ($F_B$) [8]

from either cytochrome c6 or plastocyanin to the terminal electron acceptor at its stromal side through a series of redox reactions a long Electron transfer chains. The crystal structure of a trimeric cyanobacterial PSI is resolved at atomic resolution of 2.5 Å [12], where each monomer consists of about 12 polypeptide chains (PSaA-PSaX) (Figure 1.a. and 1.b.).

There are three highly conserved chains in PS I PsaA, PsaB and PsaC[13]. The first two chains form the heterodimeric core, which non-covalently bound most of the antenna pigments, redox cofactors employed in the Electron transfer chains ETCs and the While interpolypeptide iron-sulfer cluster $F_X$[14,15]. PsaC comprises two iron-sulfur clusters $F_A$ and $F_B$, and it form, with PsaE and PsaD, the stromal hump providing a docking site for protein soluble ferredoxin[16,17] (Figure. 1.a.). Cofators employed in the ETCs are a chlorophyll (a) dimer P700, two pair of chloropyll a molecules $A/A_0$ and two phylloquinones $A_1$. These cofactors are arranged in two nearly symmetric branches A and B, from P700 at the luminal side to $F_X$ at the PsaA and PsaB interface followed by the two terminal iron-sulfur clusters $F_A$ and $F_B$, (Figure 1.c.) [8,18,19].

Upon photo-excitation of a primary electron donor P700, an electron will transfer to the primary electron acceptor $A/A_0$, within ~100 $fs$[20], followed by an electron transfer to the phylloquinone molecule within 20-50 $ps$[19]. Then the electron is transferred, sequentially, to the three Iron-sulfur clusters $F_X$, $F_A$ and $F_B$ within ~1.2 $\mu s$ [19] . It was shown that the reduced $F_B$ will directly reduce a protein soluble ferredoxin (Fd), which in turn will reduce the NADP+ to NADPH in the ferredoxin-NADP+ reductase complex (FNR) [3–7,21–23]. Knowing the redox potentials of theses cofactors is crucial for understanding the primary photosynthetic processes. However, the complexity of PS I protein complex and the electrostatic nature of interactions between charged groups and among redox centers, make it difficult to assign the measured signals to a specific redox-active center. Thus, computational methods could be a complementary technique for the characterization of redox reactions.

The three iron-sulfur clusters in PS I are 4Fe-4S clusters, which is a distorted cube of 4 Iron atoms linked by four bridging sulfur atoms and ligated by four cysteine ligands[8]. The PsaC polypeptide chain provides the cysteine ligands to both clusters $F_A$ and $F_B$; C53, C50, C20 and C47 for $F_A$ and C10, C57, C13 and C16 for $F_B$. While the $F_X$ cluster is ligated by four cysteines: two from PsaA chain (C578 and C587) and two from PsaB chain (C565 and C574). They are mainly distinguished by their low temperature EPR spectrum[24,25]. In PS I, $F_X$, $F_A$ and $F_B$ are known as low potential [4Fe-4S] clusters employ the $2^+/1^+$ redox couple[26–28]. In its oxidized state low-potential [4Fe-4S] cluster has two ferric and two ferrous Fe atoms and possess a total spin S= 0. While in its reduced state there are one ferric and three ferrous Fe atoms with total spin S=1/2. This is due to the paramagnetic pairing between an equal- valence pair $Fe^{+2}-Fe^{+2}$ and a mixed-valence pair $Fe^{+2.5}-Fe^{+2.5}$ [8].

In PS I, the redox potentials of 4Fe-4S clusters varies in a wide range from -730 to -440[19]. Where low-temperature Electron Paramagnetic Resonance EPR spectroscopy studies had showed that the midpoint potentials are -705 ± 15 , -530 and -580 mV for $F_X$, $F_A$ and $F_B$ respectively[8,19]. However, other studies suggesting that the midpoint potentials of these clusters would be positively shifted ,at room temperature[29–32]. Here, we report the calculated relative midpoint potential of [4Fe-4S] clusters $F_X$, $F_A$, and $F_B$, using Multi-Conformer

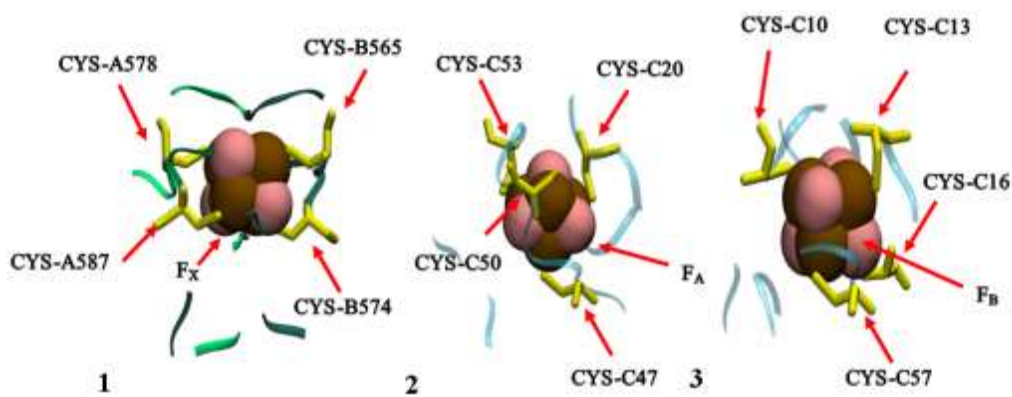

**Figure 2.** Structural models used in this study. **1, 2** and **3.** Are the iron-sulfur clusters in PS I surrounded by nearby aminoacids (~10 Å) from PsaA/PsaB and PsaC subunits. Where the letters A, B and C refers to the subunits PsaA, PsaB and PsaC, respectively. **1.** The Interpolypeptide 4Fe-4S cluster $F_X$ and the surrounding aminoacids from both protein domains PsaA and PsaB. **2. and 3.** The stromal iron-sulfur cluster $F_A$ and $F_B$, respectively, surrounded by near residues from PsaC subunit.

Continuum Electrostatics (MCCE) [33–35]. In addition, we provide an insight on the conformational changes and the interactions that induce the differences in the redox potential of the three [4Fe-4S] clusters from the classical electrostatics' perspective and their implication on the electron transfer reaction.

**Materials and methods.**

**Structural model.** Initial coordinates are obtained from the crystal structure of Thermosynechococcus elongatus (PDB code: 1JB0[12]), at resolution 2.5 Å. Structures for [4Fe-4S] clusters $F_X$, $F_A$ and $F_B$, surrounded by ~10 Å nearby residues, as shown in Figure. 2, are extracted from the crystal structure and optimized using DFT/B3LYP level of theory, with LANL2DZ basis sets[36] for Fe metal centers and 6-31G* basis set for other atoms, using Gaussian09 package[37]. The [4Fe-4S] core is set to the reduced state with total spin S= ½ using the broken symmetry wavefunction[38].

**Multi-conformer Continuum Electrostatics (MCCE) Calculations.** MCCE generates different conformers for all amino acid residues and cofactors. These conformers undergo a preselection process, which discards conformers that experience vdW clashes.[34] All crystallographic water molecules and solvated ions are stripped off and replaced with a continuum dielectric medium. The electrostatic potential of the protein is calculated by solving Poisson-Boltzmann equation[39] using DelPhi.[40] In this calculation, the surrounding solvent (water) was assigned a dielectric constant of $\epsilon = 80$, and $\epsilon = 4$ for protein.[41] The partial charges and radii used for amino acids in MCCE calculations are taken from the PARSE charges.[42] The probe radius for placing water is 1.4 Å and 0.15 M salt concentration is used. For 4Fe-4S clusters, each Fe ion, bridging sulfurs S and each ligand as separate fragments with an integer charge, which are interacting with each other through electrostatic and Lennard-Jones potenials[43].

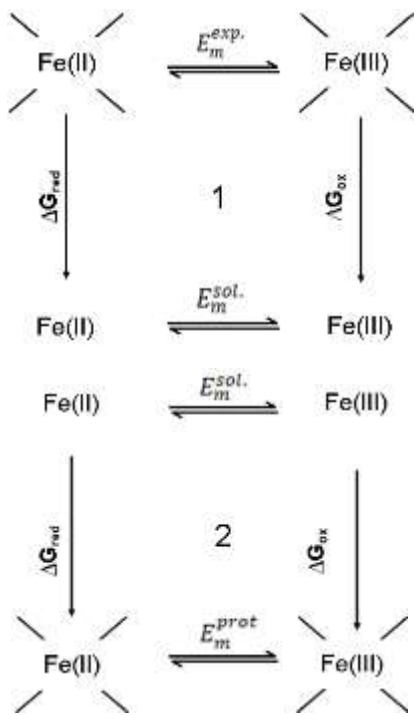

**Figure 3.** Thermodynamic cycle for the redox reaction $Fe^{+2} \rightleftharpoons Fe^{+3} + e^{-1}$. Where $E_m^{exp}$ is the midpoint potential determined in experiment, $E_m^{sol}$ is the midpoint potential in reference medium and $E_m^{prot}$ is the midpoint potential in situ calculated by MCCE.

The Fe atoms has formal charges of +2 or +3, while, each bridging sulfur atom has a charge of -2.

For each conformer i, DelPhi calculates different energy terms, the polar interaction energy ( $\Delta G_{pol,i}$ ), desolvation energy $\Delta\Delta G_{rxn,i}$, and pairwise electrostatic and Lennard-Jones interactions with other conformers j ( $\Delta G_{ij}$ ). For M conformers, the $\Delta\Delta G_{rxn,i}$ and $\Delta G_{pol,i}$ energy terms will be collected into two matrices with M rows while the $\Delta G_{ij}$ energy term will be collected into M×M matrix[35]. A single protein microstate $x$ is defined by choosing one conformer for each cofactors and residues. Therefore, number of possible microstates of the system is very high. As a final step, MCCE uses Monte Carlo sampling to compute the probability of occurrence for each conformer in the Boltzmann distribution for a given parameters pH and electron concentration ($E_h$) [35,44].

The total energy of each microstate $G_x$ with M conformers is the sum of electrostatic and non-electrostatic energies and it is computed according to the equation below [45,46]:

$$\Delta G_x = \sum_{i=1}^{M} \delta_{x,i} \left[ \left(2.3 m_i k_b T (pH - pK_{sol,i}) \right. \right.$$
$$+ n_i F(E_h - E_{m,sol,i}))$$
$$+ \left(\Delta\Delta G_{rxn,i} + \Delta G_{pol,i}\right)$$
$$\left. + \sum_{j \neq i}^{M} \delta_{x,j} \Delta G_{ij} \right] \quad (1)$$

Where $\delta_{x,i}$ is equal to 0 if microstate $x$, lacks conformer $i$ and 1 otherwise. While $m_i$ takes the values 0, 1 and -1 for neutral, bases and acid conformers, respectively. $n_i$ is the number of electrons transferred during redox reactions. $pK_{a,sol,i}$ and $E_{m,sol,i}$ are the reference $pK_a$ and $E_m$ for i[th] group in the reference dielectric medium (e.g. water). F is the faraday constant, while $K_b$ is the Boltzmann constant and T is temperature (298 K in our calculations). $\Delta\Delta G_{rxn,i}$ is the desolvation energy of moving conformer $i$ from solution to its position in protein. $\Delta G_{ij}$ is the pair-wise interaction between different conformers $i$ and $j$. While $\Delta G_{pol,i}$ is the pair-wise interaction of conformer $i$ with other groups with zero conformational degrees of freedom (e.g. Backbone atoms).

The reference solution $E_{m,sol}$ for Fe ions are obtained according to the thermodynamic cycles shown in Figure. 3. The experimental redox potential $E_{m,exp}$ of F$_A$ (440 mV vs SHE) was used to obtain the reference solution $E_{m,sol}$ for Fe$^{+3/+2}$ redox couple (-780 mV). Which was used to calculate the redox potential of the other clusters F$_B$ and F$_X$ in protein[43].

**Mean Field Energy (MFE) analysis.** MCCE determines the in-situ midpoint potential $E_m$ of the redox centers as shifted by the protein environment. This shift is due to the loss in the reaction field energy $\Delta\Delta G_{rxn}$ and other electrostatic interactions. Mean field energy analysis (MFE) allows decomposition of these energetic terms to determine what factors yield the reported midpoint potentials in protein, Eq. 2. [47],

$$nFE_{m,MFE} = nFE_{m,sol} + \Delta G_{bkbn} + \Delta\Delta G_{rxn} + \Delta G_{res}^{MFE} \quad (2)$$

Where $\Delta G_{bkbn}$ is the electrostatic and non-electrostatic interactions of the redox cofactor with the backbone atoms of protein and $\Delta G_{res}^{MFE}$ is mean-field electrostatic interaction between the redox cofactor and the average occupancy of conformers of all other residues in the protein in the Boltzmann distribution at each $E_h$ [47]. Other terms are same as shown in Eq.1.

**Results and discussion.**

Molecular structures for [4Fe-4S] clusters in PS I had been investigated by extended X-ray absorption fine structure (EXAFS), which revealed two peaks at ~2.27 Å and ~2.7 Å, which are attributed to the backscattering from sulfur and iron atoms, respectively.[48–51] The results of geometry optimization of three extracted structures with total spin S = ½ and with [4Fe-4S] in their reduced state, are reported in Table 1.a.

Our calculated Fe-S (bridging sulfur atoms), Fe-SG (Organic sulfur atoms) and Fe-Fe bond distances are shown, generally, to be longer than the XRD[12] and EXAFs reported distances Table 1.a and b, respectively.

**The midpoint potentials ($E_m$) of F$_X$, F$_B$ and F$_A$ at pH 10.** In our calculations, we considered the oxidation potential of the 2$^{nd}$ oxidized Fe ion as the oxidation potential of the cluster from [4Fe-4S]$^{+1}$ to [4Fe-4S]$^{+2}$. The measured $E_m$ values of F$_X$, F$_A$, and F$_B$ are reported in Table. 2. For F$_X$, $E_m$ is -796 mV which is ~91-146 mV more negative than experimental values [8,52,53]. While measured $E_m$s for F$_A$ and F$_B$ are -454 and -545 mV, respectively. Which lies within the range of experimentally determined values [19,54,55]. Our results are shown to agree with the experimental values within the error range of the method [35,56,57]. To better understand the effect of ligands and other residues in the model structures on the calculated $E_m$s, Mean Field Energy (MFE) analysis is performed for each 4Fe-4S cluster at its calculated $E_m$ to determine the different factors contributing to the stabilization of ionization state of clusters in protein, Eq. 2.

Table 1.a. Bond distances of 4Fe-4S Clusters from XRD experiments and DFT geometry optimization

|  | DFT | | | XRD | | |
| --- | --- | --- | --- | --- | --- | --- |
|  | $F_X$ | $F_A$ | $F_B$ | $F_X$ | $F_A$ | $F_B$ |
| Fe-S (Å) | 2.32 | 2.32 | 2.34 | 2.3(×1) | 2.3(×7) | 2.3(×12) |
|  | 2.35 | 2.37 | 2.37 | 2.2(×1) | 2.2(×4) |  |
|  | 2.39 | 2.38 | 2.38 |  | 2.4(×1) |  |
|  | 2.45 | 2.39 | 2.4 |  |  |  |
|  | **2.46** | 2.4 | 2.4 |  |  |  |
|  | 2.47 | 2.44 | 2.4 |  |  |  |
|  | 2.49 | 2.46 | 2.44 |  |  |  |
|  | 2.52 | **2.46** | **2.44** |  |  |  |
|  | 2.52 | 2.47 | **2.49** |  |  |  |
|  | **2.44** | 2.48 | 2.49 |  |  |  |
|  | **2.44** | **2.52** | **2.5** |  |  |  |
|  | 2.44 | 2.57 | 2.53 |  |  |  |
| Fe-SG (Å) | 2.36 | 2.49 | 2.39 | 2.4(×2) | 2.4(×1) | 2.4(×2) |
|  | 2.37 | 2.35 | 2.4 | 2.2(×1) | 2.3(×1) | 2.3(×2) |
|  | **2.35** | **2.34** | **2.35** | 2.3(×1) |  |  |
|  | 2.34 | 2.34 | 2.36 |  |  |  |
| **Avg.** | **2.42** | **2.45** | **2.45** |  |  |  |
| Fe-Fe (Å) | 2.96 | 3.16 | 3.05 | 2.7(×6) | 2.7(×4) | 2.7(×6) |
|  | 2.97 | 3.02 | 3.14 |  |  |  |
|  | 3.04 | 3.19 | 2.83 |  |  |  |
|  | 3.18 | 2.95 | 3.15 |  |  |  |
|  | 3.29 | 3.2 | 3.1 |  |  |  |
|  | 3.15 | 2.97 | 3.02 |  |  |  |

a; bold values are the distances between the 2nd oxidized Fe ion and the four sulfur ligands (3 bridging sulfurs and one from cysteine), while the Avg. is the average over these distances for each 4Fe-4S cluster, see Figure 4.

Table 1.b. Bond distances determined from EXAFS studies

| EXAFS (Å) | |
| --- | --- |
| Fe-S | Fe-Fe |
| 2.27 | 2.7 |

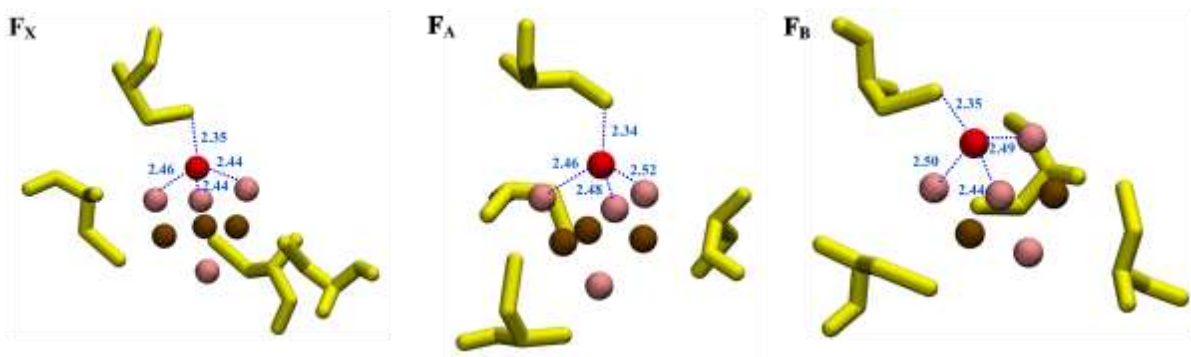

**Figure 4.** The structure of optimized $F_X$, $F_A$, and $F_B$ redox centers, showing distances between 2$^{nd}$ oxidized Fe ion and the four ligating sulfurs (three bridging sulfurs and one sulfur from cysteine). Red spheres are the 2$^{nd}$ oxidized Fe ion, while brown spheres are the other Fe ions in the cluster. pink spheres are bridging sulfurs and finally yellow sticks are Cysteine ligands.

**Table 2**. Calculated Midpoint potential for redox couples +2/+1 [in units of mV]

|  | Cal. $E_m$s | Exp. $E_m$s |
|---|---|---|
| $F_A$ | **-453** | **-440**[j], -530[i], -500[m] |
| $F_B$ | -546 | -465[j], **-580**[i], -550[m] |
| $F_X$ | -796 | -650[m], **-705**[k], -670[l] |

The bold is the Em value used as a reference.

[i]ref(24,58), [j]ref(55), [k]ref(52), [l]ref(31), [m]ref(59)

Results from the MFE analysis is reported in Table 3. The desolvation energy term $\Delta\Delta G_{rxn}$ was shown to be always positive and unfavorable energy term. It destabilizes the ionization state of the iron-sulfur clusters in structures **1**, **2**, and **3** by ~69, ~63 and ~57 Kcal/mol, respectively. This unfavorable interaction is, nearly, compensated by the electrostatic interactions with the surrounding residues $\Delta G_{resd}$ in structures **1**, **2**, and **3**, to be -73, -62, and -59 Kcal/mol, respectively. Moreover, interactions with the backbone $\Delta G_{bkbn}$ disfavor the oxidized form of Fe ion. For $F_A$ and $F_B$ this effect is shown to be, significantly, ~ 2-fold more than that in $F_X$. By further breaking down the contribution from different residues and ligands (see Table 4.), it is shown that stabilization of ionization state of 4Fe-4S clusters is mainly controlled by the classical electrostatic interactions between Fe ions and both of bridging sulfurs and cysteine ligands. Oppositely, the electrostatic interaction with positively charged residues and other Fe ions are shown to destabilize the oxidized form of Fe ion. The total electrostatic interaction energy within the clusters are shown to be -47.1, -38.48 and -34.31 Kcal/mol for $F_X$, $F_A$, and $F_B$, respectively. The low potential of $F_X$ is shown to be due to the backbone and residue sidechains contributions. Moreover, the distances between ligating sulfurs and the 2$^{nd}$ oxidized Fe ion were found to be, on average, 2.42 Å for $F_X$, and ~2.45 Å for $F_A$ and $F_B$. Which could explain the higher effect of sulfurs in $F_X$ for shifting the redox potential.

Redox potential of iron-sulfur clusters in PS I were calculated previously, by the work of Torres et al.[60]. They reported the values of $E_m$'s for $F_X$, $F_A$ and $F_B$ to be -980, -510, and -710 mV, respectively.

**Table 3.** Energy terms that contributes to the shift of the redox potential in protein. These terms are shown to be the desolvation energy term $\Delta\Delta G_{rxn}$, backbone contribution $\Delta G_{bkbn}$, and pairwise interaction with sidechains $\Delta G_{resd}$. [energies are in units of Kcal/mol]

| $\Delta G^a$ | $F_X$ | $F_A$ | $F_B$ |
|---|---|---|---|
| $\Delta G_{bkbn}$ | 3.94 | 8.10 | 7.56 |
| $\Delta\Delta G_{rxn}$ | 69.28 | 63.09 | 57.83 |
| $\Delta G_{resd}$ | -73.34 | -61.87 | -59.76 |

Where the $E_m^{sol}$ was obtained by correcting the ionization potential calculated by gas-phase DFT with the solvation effects and referencing the calculated potential to the standard hydrogen electrode ($\Delta SHE = -4.5\ eV$). Torres et al. employed a model with three dielectric regions, the continuum solvent ($\varepsilon_{wat} = 80$), the protein ($\varepsilon_{prot} = 4$) and $\varepsilon = 1$ for the redox site to reflect the little screening effect of protein due to hydrogen bonding in the vicinity of the clusters. In their paper, Ptushenko et al.[61] argued the implausibility of the proposed 3 dielectric regions model by Torres due to the overestimation of the amide field in the vicinity of clusters. Which lead to a negatively deviated midpoint potential from experimental values by 275 to 330 mV for $F_X$ and 130 to 245 mV for $F_B$.

Also in the work of Ptushenko and coworkers[61], they calculated midpoint potentials for all redox cofactors in PS I including the three [4Fe-4S] clusters. Their reported values are -654, -481 and -585 mV for $F_X$, $F_A$ and $F_B$, respectively. In their calculations they employed the semi-continuum electrostatic model. Where, two dielectric constants for proteins were used, the optical dielectric constant ($\varepsilon_o = 2.5$) for pre-existing permanent charges and a static dielectric constant ($\varepsilon_s = 4$) for charges formed due to formation of ions in protein upon ionization reaction. In their calculations, Torres and Ptushenko included all protein subunits and other prosthetic groups in PS I complex. Although, we only included residues within ~10 Å surrounding each cluster, our results showed a high correlation to the experimentally determined midpoint potentials. Our results suggest that the contribution of distant residues might be minimal compared to the effect of interaction with negatively charged sulfur ligands.

**CONCLUSION.**

We have documented for the first time the redox potential calculation of iron-sulfur clusters in PS I using the MCCE Model. Good agreement between calculated and experimental midpoint potentials is obtained for the +1/+2 redox couple in 4Fe-4S clusters. Our calculations showed that the stabilization of the oxidized state of the 4Fe-4S clusters in protein is mainly due to the pairwise interaction with residues side chains. The fact that $F_x$ is an unusual low-potential cluster may be attributed to the bond length between the oxidized Fe ion and sulfur ligands, which is shown to be shorter than that for both of $F_A$ and $F_B$. Also, interactions with backbone atoms are shown to be least for $F_X$.

Table 4. Electrostatic interaction between iron-sulfur clusters and the surrounding residues [Energies are in units of Kcal/mol][a]

|  | $F_X$ | | | | $F_A$ | | $F_B$ | |
|---|---|---|---|---|---|---|---|---|
|  | PsaA | | PsaB | | PsaC | | PsaC | |
|  | Residues | Energy | Residues | Energy | Residues | Energy | Residues | Energy |
| Surrounding residues | S3 | -16.76 | C565 | -5.51 | S2 | -14.64 | S2 | -14.59 |
|  | C587 | -4.16 | C574 | -3.56 | C50 | -4.07 | C57 | -4.12 |
|  | T586 | 0.37 | T573 | 0.33 | C47 | -3.66 | C13 | -3.9 |
|  | R728 | 2 | R712 | 1.04 | C53 | -3.16 | C10 | -2.76 |
|  |  |  |  |  | T22 | 0.09 | S63 | -0.16 |
|  |  |  |  |  | K51 | 0.8 | T14 | -0.07 |
|  |  |  |  |  | R52 | 1.23 | M27 | -0.06 |
|  |  |  |  |  |  |  | T59 | 0.03 |
|  |  |  |  |  |  |  | Q15 | 0.19 |
| 4Fe-4S | Fe1 | 28.39 |  |  | Fe1 | 27.41 | Fe4 | 24.56 |
|  | Fe2 | 29.33 |  |  | Fe3 | 27.65 | Fe3 | 26.12 |
|  | Fe4 | 50.04 |  |  | Fe4 | 41.59 | Fe1 | 42.67 |
|  | **S2** | **-44.41** |  |  | **S4** | **-40.61** | **S4** | **-37.85** |
|  | **S1** | **-44.27** |  |  | **S3** | **-39.55** | **S3** | **-36.6** |
|  | **S4** | **-43.96** |  |  | **S1** | **-35.2** | **S1** | **-36.33** |
|  | **C578** | **-22.22** |  |  | **C20** | **-19.77** | **C16** | **-16.88** |

The bold is the ligands for the 2nd oxidized Fe ion.

[a] Residues are represented by the single letter code, while (S1, S2, S3 and S4) and (Fe1, Fe2, Fe3 and Fe4) are the bridging sulfur ions and Fe ions in 4Fe-4S clusters, respectively.

**REFERENCES.**